\documentstyle[preprint,aps,epsfig]{revtex}

\newcommand{\be}{\begin{equation}}
\newcommand{\ee}{\end{equation}}
\newcommand{\bear}{\begin{eqnarray}}
\newcommand{\ear}{\end{eqnarray}}
\newcommand{\la}{\label}

\begin{document}

ADP-96-28/T227  \hspace*{6cm} August 1996

\vspace{2cm}
\centerline{\large The Flavor Asymmetry of the Nucleon Sea}
\vspace{.4cm}
\centerline{F.M. Steffens\footnote{Current address: Instituto de Fisica - UFRGS,
Caixa Postal 15051, Porto Alegre - RS, CEP:91501-970, Brazil. Email: fsteffen@sofia.if.ufrgs.br} and A.W. Thomas\footnote{athomas@physics.adelaide.edu.au}}
\vspace{1cm}
\centerline{Department of Physics and Mathematical Physics,}
\centerline{and Institute for Theoretical Physics,}
\centerline{University of Adelaide,}
\centerline{Adelaide, S.A. 5005, Australia}
\vspace{.3cm}
\begin{abstract}
We re-examine the effects of anti-symmetry on the anti-quarks in the
nucleon sea arising from gluon exchange and pion exchange between confined
quarks. While the effect is primarily to suppress anti-down relative to
anti-up quarks, this is numerically insignificant for the pion terms.

\end{abstract}

\section{Introduction}

Since the precise measurement of the Gottfried sum rule
\cite{gottfried67} by the New Muon Collaboration (NMC) \cite{nmc91} in
1991, the possibility that $\overline u (x) \neq \overline d (x)$ in the
proton has been extensively studied. The reason for this attention is that, in
the quark-parton model, the Gottfried sum rule is expressed as:

\be
S_G = \int_0^1 [F_2^{p}(x) - F_2^{n}(x)]\frac{dx}{x} = \frac{1}{3} 
+ \frac{2}{3}\int_0^1 [\overline{u} (x) - \overline d (x)],
\label{0}
\ee
where charge symmetry between the proton and the neutron was used.
If SU(2) flavor symmetry, $\overline u = \overline d$, is also assumed, 
$S_G$ is reduced to 1/3.

In order to test the Gottfried sum rule experimentally, the NMC \cite{nmc91} performed 
deep inelastic muon scattering on hydrogen and deuterium and measured
the deuteron/proton cross-section ratio. To extract the proton
and neutron structure functions, they used the simple relation
$F_2^p (x) - F_2^n (x) = 2F_2^d (x) (1 - F_2^n (x)/ F_2^p (x))/
(1 + F_2^n (x) / F_2^p (x))$, with $ F_2^n (x) / F_2^p (x)= 2F_2^d (x)
 / F_2^p (x) -1$. A parametrization \cite{nmc91,nmc92} for $F_2^p (x)$
based on NMC, SLAC and BCDMS data was used.  
The NMC reported the following value for the Gottfried sum
rule:

\begin{eqnarray}
S_G(0.004\leq x \leq 0.8) &=& 0.221\pm 0.008(stat)\pm 0.019(syst), 
\nonumber \\*
S_G &=& 0.235\pm 0.026 ,
\label{00}
\end{eqnarray}
where the contribution for $x>0.8$ was estimated using a smooth
extrapolation of $F_2^n / F_2^p$ and for the region $x<0.004$ 
a Regge-like behavior ($ax^b$ with $a=0.2\pm 0.03$ and
$b=0.59\pm 0.06$) was assumed. It is worth noticing that in the extraction of
$F_2^n / F_2^p$, target mass corrections and nuclear effects were
not included, nor were higher twist corrections included
in the sum rule as a whole. However, a detailed examination of binding,
shadowing and meson exchange (anti-shadowing)
corrections \cite{wally93} does not greatly change 
their conclusions.

The experimental result from the NMC yields:

\be
\int_0^1 dx [\overline u(x) - \overline d (x) ]= -0.147 \pm 0.039.
\label{000}
\ee 

Theoretically, there has been some 
indication for $\overline u \neq \overline d$ 
since the work of Feynman and Field \cite{feynman77}, where the Pauli principle
was invoked to justify that $u\overline u$ pair creation is suppressed relative
to $d\overline d$. Later, Ito et al. \cite{ito} measured continuum dimuon production and
determined the sea quark distribution 
in the context of the Drell-Yan description
of dilepton production. From an analysis of the 
logarithmic derivative of the measured
cross section, they inferred that the Drell-Yan model, assuming a symmetric
sea, underestimated the measured slope. This fact was interpreted 
as an indication of broken symmetry in the sea, 
with $\overline u < \overline d$.
Somewhat later, Thomas \cite{tony83}, following Sullivan's suggestion 
\cite{sullivan} that the pion cloud can contribute to the nucleon 
structure function, realized that pion dressing of the nucleon
naturally predicts an excess of $\overline d$ over $\overline u$. This is
because, for instance, the proton has a greater probability to 
emit a $\pi^+$ than a $\pi^0$ or a $\pi^-$. In fact, this observation was
used \cite{tony84} to interpret the Ito et al. results on dilepton production.
Since then, the idea of pions producing an 
asymmetric sea has been widely explored 
\cite{find} as a possible interpretation of the NMC results. 
Further theoretical evidence for asymmetry in the sea was found in
the calculation of quark distributions of Signal and Thomas \cite{signal89}. 
The matrix elements of the quark distribution involve intermediate
states where the photon scatters on a valence quark and also intermediate 
states where the photon oscillates into a $q\overline q$ pair and a quark or 
antiquark is inserted in the nucleon. Again, following the idea of Field and 
Feynman, it is easier to insert a $d$ than a $u$ quark into the proton. 

Even with all this evidence, until the NMC experiment, the nucleon  
continued to be seen as having a symmetric sea. There is a simple explanation
for this. The whole set of data available could be described by 
parametrizations of parton distributions where it was assumed that
$\overline u(x) = \overline d(x)$, with the price of a slightly odd
behaviour of the valence quark distribution when $x\rightarrow 0$ \cite{mrs93a}.
In this line of thought, it was also proposed that the NMC result 
could still be reproduced with flavor and charge symmetry but 
with a modified behaviour for $F_2 (x)$ at small $x$, in such way that
the integral $(F_2^p (x) - F_2^n (x))/x$ saturates the naive 
expectation 1/3 \cite{mrs90}. However, the NA51 collaboration \cite{na51}
recently carried out an experiment proposed by Ellis and Stirling \cite{ellis91}
to distinguish between a symmetry breaking effect
and an odd small $x$ behaviour of the parton distributions. 
The NA51 results indicate
a strong flavor symmetry breaking at $x=0.18$, thus reinforcing the
idea that the sea is indeed flavor asymmetric. In a recent reanalysis 
\cite{steffens96} of
the NA51 result, this conclusion was reinforced, although the
exact size of the flavor asymmetry is dependent on the extent of
the charge symmetry breaking between the proton and the neutron. Finally,
calculations based on meson cloud models \cite{harald} are also able to produce
an asymmetry compatible with what is measured by the NA51 collaboration. 

We then see that the Pauli principle and pion dressing of the 
nucleon are two powerful tools to understand flavor symmetry breaking.
If a bare proton is seen as having three valence quarks, 
two up and one down, then
all the antiquarks have their origin in gluonic and/or 
pionic effects. For instance,
pair creation through gluon emission of one of the valence quarks will
produce a intermediate state of four quarks and one antiquark. The flavor 
of the antiquark is certainly restricted by the valence structure of the proton
via the Pauli principle. However, one should also account for 
the antisymmetrization between the various quarks. In this case, the valence
structure will also interfere in the number of antiquarks created.
In order to understand the interplay of these effects, we study in some 
detail the relation between the 
Pauli principle and quark antisymmetrization in pair creation inside 
a nucleon. 

In the pionic case the same reasoning applies. Of course, the pion is
complicated by the fact that it is a pseudo-Goldstone boson. Thus its
internal structure will be quite complicated and the probability to find
just a $q-\bar{q}$ pair with the quark 
in a 1-s state, would be expected to low. As
anti-symmetrization with the valence quarks in the nucleon will only
effect quarks in the 1-s state (of the nucleon bag) we expect that
anti-symmetry should have a rather small effect on pion loops.
Nevertheless, it is important to quantify this prejudice and that is the
main purpose of the present calculation.
 
\section{The Pauli Principle in the Proton Sea}

The most natural idea to account for the observed discrepancy between 
theory and
experiment is to invoke the Pauli principle. Field and Feynman 
\cite{feynman77} were the first ones to use this idea: if the
proton is composed of two valence $u$ quarks and one valence
$d$ quark then the creation of a quark - antiquark pair 
through gluon emission will tend to give more $d\overline{d}$ pairs
than $u\overline{u}$ pairs. This is easy to understand
because there are five empty states 
for the $d$ quark and four for the $u$ quark. Although this
idea is attractive and essentially correct, there is one other effect
to consider. We will see that we also have to 
consider graphs containing interference between the sea quarks 
generated in the gluon emission and the remaining quarks in the nucleon. 
This effect will, in fact, hide the excess of $\overline{d}$ 
over $\overline{u}$ due to the Pauli principle. 

To illustrate our discussion, we begin by reviewing the pair creation through
gluon exchange, following the calculations of Donoghue and
Golowich \cite{donoghue}. The assumption made is that the bare proton is 
composed of two valence $u$ quarks, one valence
$d$ quark and its color, flavor and spin wave function is given by:

\begin{equation}
|p\rangle_0 = \frac{1}{\sqrt{18}}\epsilon^{\alpha \beta \gamma}[b^{\dagger} (u,\uparrow,\alpha)
b^{\dagger} (d,\downarrow,\beta) - b^{\dagger} (u,\downarrow,\alpha)
b^{\dagger} (d,\uparrow,\beta)] b^{\dagger} (u, \uparrow, \gamma) |0\rangle .
\label{1}
\end{equation}
The sea quarks are generated through a quark gluon interaction
given by:

\begin{equation}
H_I (x)=g\overline{\psi}(x)\gamma^{\mu}\frac{\lambda^a}{2}\psi (x){\it A}^{a}_{\mu} (x),
\label{2}
\end{equation}
where $g$ is the coupling constant, $A^a$ are the gluon fields and $\psi$ the 
quark field. We are not going to worry about the form of the spatial part
of these operators but will concentrate only on 
the color, spin and flavor part of the proton dressed with a quark - antiquark 
pair:

\begin{equation}
|p\rangle = |p\rangle_0 +  \frac{1}{E_0 - H_0}H_I 
\frac{1}{E_0 - H_0}H_I |p\rangle_0 + ...,
\label{3}
\end{equation}
with $H_0$ the free Hamiltonian. (Note that we have omitted the term corresponding to 
a single gluon with the three valence quarks.) 

The vector coupling between quarks and gluons allows for vertices
where the quark that emits the gluon either changes its spin or not.
Another feature of the pair creation process is that, since a particle
has opposite intrinsic parity to an antiparticle, at least one quark (or 
the antiquark) or the two quarks and the antiquark have to be in a state of odd parity
in order to conserve the parity of the proton. 
The proton wave function is then written as:

\begin{equation}
|p\rangle = |p\rangle_0 + C_s |\psi_s\rangle + C_v |\psi_v\rangle ,
\label{4}
\end{equation}
with $|\psi_v\rangle$ the wave function for the case where the
quark that emits the gluon can change its spin and $|\psi_s\rangle$ 
the wave function for 
the case that the quark emitting the gluon does not change its 
spin. The factors $C_s$ and $C_v$ depend on the particular form 
for the spatial wave functions, according to Eq. (\ref{3}).
We then have:
\bear
\langle p|p\rangle &=& 1 + |C_s|^2 \langle \psi_s|\psi_s\rangle
+ |C_v|^2 \langle \psi_v|\psi_v\rangle  \nonumber \\*
     &+& C_s^* C_v \langle \psi_s | \psi_v \rangle + C_v^* C_s \langle \psi_v | \psi_s \rangle . 
\label{5}
\ear
The wave function involved in the vector coupling is calculated 
from Eqs. (\ref{2}) and (\ref{3}):

\bear
|\psi_v\rangle &=& (-1)^{\tilde{n} -1/2}b^{\dagger}(s,n,\rho)
\sigma^{l}_{n\tilde{n}}\lambda^a_{\rho\tilde{\rho}}
d^{\dagger}(\overline s , -\tilde{n},\tilde{\rho}) \nonumber \\*
             && b^{\dagger}(v,m,\delta)\sigma^{l}_{m \tilde{m}}
\lambda^a_{\delta \tilde{\delta}} b(\tilde{v},\tilde{m},\tilde{\delta})|p\rangle_0 
\nonumber \\*
            &=&\frac{(-1)^{\tilde{n} -1/2}}{\sqrt{18}}\sigma^{l}_{n\tilde{n}}
\sigma^{l}_{m \tilde{m}}\lambda^a_{\rho\tilde{\rho}}\lambda^a_{\delta \tilde{\delta}}
\nonumber \\*
  &&\{\epsilon^{\alpha\beta\gamma}(
\delta_{\tilde{v}u}\delta_{\tilde{\delta}\alpha}A^{\dagger}
- \delta_{\tilde{v}d}\delta_{\tilde{\delta}\beta} B^{\dagger}
+ \delta_{\tilde{v}u}\delta_{\tilde{\delta}\gamma} C^{\dagger})|0\rangle \},
\label{6}
\ear
where $\sigma^l$ are the Pauli spin matrices, $\lambda^a$ are the Gell-Mann
color matrices and the spin and flavor indices are summed. We also 
have:

\bear
A^{\dagger} &=&b^{\dagger}(s,n,\rho)d^{\dagger}(\overline s , -\tilde{n},\tilde{\rho})
[b^{\dagger}(v,m,\delta)b^{\dagger}(d,\downarrow,\beta)\delta_{\tilde{m}\uparrow}
-b^{\dagger}(v,m,\delta)b^{\dagger}(d,\uparrow,\beta)\delta_{\tilde{m}\downarrow}]
b^{\dagger}(u,\uparrow,\gamma) \nonumber \\*
B^{\dagger} &=&b^{\dagger}(s,n,\rho)d^{\dagger}(\overline s , -\tilde{n},\tilde{\rho})
[b^{\dagger}(v,m,\delta)b^{\dagger}(u,\uparrow,\alpha)\delta_{\tilde{m}\downarrow}
-b^{\dagger}(v,m,\delta)b^{\dagger}(u,\downarrow,\alpha)\delta_{\tilde{m}\uparrow}]
b^{\dagger}(u,\uparrow,\gamma) \nonumber \\*
C^{\dagger} &=&b^{\dagger}(s,n,\rho)d^{\dagger}(\overline s , -\tilde{n},\tilde{\rho})
b^{\dagger}(v,m,\delta)\delta_{\tilde{m}\uparrow}[b^{\dagger}(u,\uparrow,\alpha)
b^{\dagger}(d,\downarrow,\beta)
-b^{\dagger}(u,\downarrow,\alpha)b^{\dagger}(d,\uparrow,\beta)].
\label{7}
\ear
For the scalar coupling  the wave function is the same except that 
the Pauli spin matrices are omitted and $n=\tilde n$, $m=\tilde m$. 

The calculation of the wave function overlap is very long and tedious.
The results are displayed in Table \ref{tb21}, where $v$ refers to the 
state of the valence quark after the gluon emission, $s$ corresponds to the state
where the quark in the sea is created and $g$ refers to the ground state.
Notice that there is no interference between the second and the third lines of
the table because of angular momentum conservation.
These calculations agree with the results of Donoghue and Golowich \cite{donoghue}.

\begin{table}
\begin{center}
\begin{tabular}{|c|c|c|c|}
State & $\langle \psi_s|\psi_s\rangle$ & $\langle \psi_v|\psi_v\rangle$ & 
$\langle \psi_s|\psi_v\rangle$ \\ \hline
v=s=g & 0 & 4608 + 1792$\delta_{su}$ - 1024$\delta_{sd}$ & 0 \\
v=g,s$\neq$g & 0 & 4608 & 0 \\
v$\neq$g, s=g & 1152 + 640$\delta_{su}$ + 320$\delta_{sd}$ &
3456 + 1280$\delta_{su}$ - 320$\delta_{sd}$ & -320$\delta_{su}$ - 640$\delta_{sd}$ \\
v=s$\neq$g & 1152 + 128$\delta_{su}$ + 64$\delta_{sd}$ &
3456 - 384$\delta_{su}$ - 192$\delta_{sd}$ & 384$\delta_{su}$ + 192$\delta_{sd}$ \\
v$\neq$s, v$\neq$g, s$\neq$g & 1152 & 3456 & 0 \\
\hline
\end{tabular}
\end{center}
\caption{Amplitudes of probabilities to find a quark in the sea for the various
possible states of the sea quark and the valence quark that 
emits the gluon. These probabilities are based only on the 
spin-flavour-colour wave function.}
\label{tb21}
\end{table}

The most striking result to be read from Table \ref{tb21} is that  
the probability to find the sea with a $u\overline u$ pair 
is bigger than the sea composed of a $d\overline d$ pair. 
This conclusion is the opposite of the experimental result collected 
by the $NMC$ \cite{nmc91} where the measured 
Gottfried sum rule is smaller than 1/3 - a result that implies
$\overline d > \overline u$.
How can we then understand that the intrinsic sea in the 
proton generated by gluons favours $u\overline u$ pairs over 
$d\overline d$ pairs? In principle, this result also contradicts the 
intuitive picture introduced by Field and Feynman \cite{feynman77},
based on the Pauli principle. For each flavor there are 
6 empty states (2 from spin and 3 from color). As the bare proton has
2 $u$ quarks and one $d$ quark, there are four available states for 
insertion of $u$ quarks and 5 available states for insertion of $d$ quarks. 
Although this reasoning is correct, one also cannot forget that 
antisymmetrization between quarks is an additional complication.
The same excess of $u$ valence quarks that prevents $u\overline u$ pair
creation in comparison with $d\overline d$ pair creation, also 
produces extra contributions because of antisymmetrization between the
$u$ quarks that does not exist for the $d$ quarks.

\begin{figure}[htb]
\vspace{8cm}
\centering{\
    \epsfig{angle=270,figure=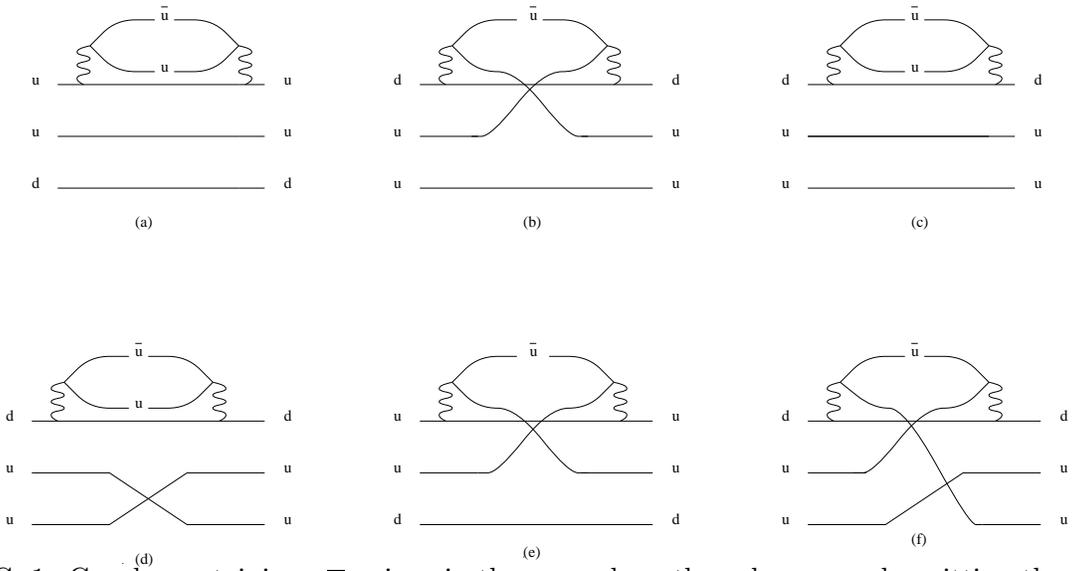,height=7.5cm}  }
\caption{Graphs containing $u\overline u$ pairs - in the case where the valence 
quark emitting the virtual gluon goes to an excited state (i.e. $s=g, v\neq g$).}
\label{figb2}
\end{figure}

\begin{figure}[htb]
\vspace{8cm}
\centering{\
    \epsfig{angle=270,figure=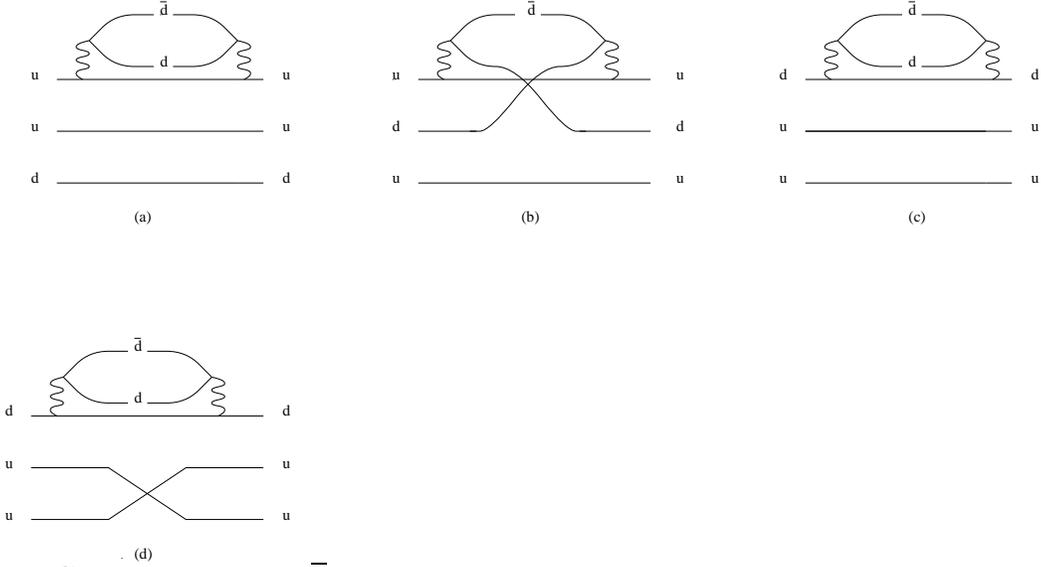,height=7.5cm}  }
\caption{Graphs containing $d\overline d$ pairs - in the case where the valence 
quark emitting the virtual gluon goes to an excited state (i.e. $s=g, v\neq g$).}
\label{figb3}
\end{figure}

To understand this effect in more detail, we consider the case 
where the quark that emits the gluon does not change its spin 
but goes to an excited state and the sea is  
created in the ground state or in a P state, according to
parity conservation. This case was chosen because it is 
the simplest, as can be seen from table \ref{tb21}. 
In Fig. \ref{figb2} we show the graphs containing $u\overline u$ pairs  
and in Fig. \ref{figb3} the graphs containing $d\overline d$ pairs.

The diagrams (a), (c) and (d) for $u\overline u$ are the analogues
of the diagrams (a), (c) and (d) for $\overline d$. Their values are
respectively the same because the created sea does not mix with the 
valence quarks and, for this reason, it is not possible to distinguish 
the sea flavor. Thus, in the case where the sea contracts with itself,
the values of these graphs are the same for any flavor. Of course, the 
heavier the quark the smaller its contribution through these diagrams,
but this is an effect related solely to mass and not to spin statistics.
The computed value is 
$\langle \psi_s|\psi_s\rangle=384$ for (a), $\langle \psi_s|\psi_s\rangle=
128$ for (c) and $\langle \psi_s|\psi_s\rangle=64$ for (d).
The diagram (b) for $\overline u$ is the analogue of diagram 
(b) for $\overline d$ and they only exist when $u\overline u$ and
$d\overline d$ pairs are created. Their values are $\langle \psi_s|\psi_s\rangle=
4\times 160/3$ for (b) of Fig. \ref{figb2} and $\langle \psi_s|\psi_s\rangle
=6\times 160/3$ for (b) of Fig. \ref{figb3}. Generally speaking,  the set of
diagrams just mentioned are the only ones which are comparable. From
them, one easily sees the Pauli principle in action and
from it an excess of $\overline d$ over $\overline u$ as expected.
However, we also have to include graphs 
(e) and (f) of Fig. \ref{figb2} for $\overline u$. They appear because 
there is an excess of $u$ valence
over $d$ valence quarks. 
If the graphs (e) and (f) are included, as they should
be, then the results change because these graphs give positive 
contributions: $\langle \psi_s|\psi_s\rangle=320$ for (e) 
and $\langle \psi_s|\psi_s\rangle=2\times 160/3$ for (f). With these
graphs included the probability to find a $\overline u$ antiquark in the proton
is bigger than the probability to find a $\overline d$ in the same proton.

One could doubt this interpretation of the relation between the
Pauli principle and antisymmetrization effects if one were to count only those 
graphs where a $d$ quark emits a gluon. In this case, where a 
$d$ quark emits a gluon,
we would have 6 empty states for a $d$ quark and only 4 
empty states for the $u$ quark (if the sea is created in the
ground state). On the other hand, from Fig. \ref{figb2}
we see that graphs (b), (c) and (d) (also graph (f) if we 
include the one coming exclusively from the excess of $u$ valence
quarks) give a larger contribution than the corresponding graphs
(c) and (d) of Fig. \ref{figb3}. As the graphs (b), (c) and (d), in principle, are not
related to the excess of $u$ valence quarks, there is an apparent
contradiction with the simple counting of states and the whole argument
of the role of the interference graphs presented by us before.
In reality, this contradiction is only apparent.

To better understand what is happening, consider a proton made of only 
one $u$ and one $d$ quark. We consider again the case where the quark 
that emits the gluon goes to an excited state.
In this case, if a $d$ quark emits a gluon, we would
have in principle 6 empty states for the insertion of a $d$ quark and $5$ 
for the insertion of a 
u quark. The possible graphs would be the analogue of (b) and (c) from
Fig. \ref{figb2} for a $u\overline u$ in the sea and the analogue 
of graph (c) from Fig. \ref{figb3} for a $d\overline d$ in the sea. 
Again, we would have more $u\overline u$ pairs than $d\overline d$ pairs
and now it is clear why that happens: this is because there is one free
valence $u$ quark that can be exchanged with the sea and there is 
no such free valence $d$ quark to be exchanged (in the case of a $d\overline d$ 
sea). The opposite situation happens when the $u$ quark emits the
gluon such that the sum of all diagrams, gluon emission from 
$u$ and $d$ valence quarks, renders an equal probability for a
$u\overline u$ and $d\overline d$ pair creation, as expected in 
a proton containing only one quark of each flavor. The lesson is that
we cannot treat the gluon emission in the proton from different 
flavors separately, and expect the Pauli principle to work free 
of any other effects.

For the combined result, the probability to find a $\overline u$
is bigger than the probability to find a  $\overline d$.
One then can say that interference terms overcome the 
naive expectation of the Pauli principle. This result is extremely important 
because it also says that if we have to antisymmetrize the sea 
quarks with valence quarks in the case where the sea quarks are
generated through pions, then the whole set of conclusions about
the importance of the pions to the Gottfried sum rule might need to 
be revised. In section III we investigate whether 
this is in fact the case.

\subsection{Discussion of the role of anti-symmetry}

Before discussing pions it is of interest to relate the findings which 
we have just reported to a similar issue in lattice QCD.
Figures 2(a), 2(c) and 2(d) are a very simple example of the
class of diagrams known in lattice QCD as disconnected insertions (DI)
\cite{leinweber96}.
While the DI's in lattice calculations include an infinite number of 
gluon exchanges, the simple example which we have just presented 
illustrates an important principal. The intermediate 4q states in 
Figs. 2(a), 2(c) and 2(d) are unphysical in the sense that their wave function
is not totally anti-symmetric. Only when the exchange process, shown 
in Fig. 2(b), is included do we have a physically meaningful result.
Within the lattice terminology the latter are connected insertions (CI).
Thus we find that neither the CI, nor the DI are physically meaningful
alone. This must be taken into account when one is trying to interpret
the physical mechanisms behind lattice results for flavour breaking 
in the nucleon sea, the $\sigma$ commutator and so on. 

\section{Antisymmetrization of the Quarks of the Pion}

We now wish to investigate the effect of anti-symmetrization on the
internal structure of the pion. As already noted in the introduction,
the pion (as a pseudo-Goldstone boson) is expected to have a complicated
internal structure. The effect of anti-symmetrization on this internal
structure can only effect components of the wavefunction where a quark
is in the 1-s state of the nucleon bag. We consider the case where the
pion couples to just one $q-\bar{q}$ pair with the extra quark in the
1-s state. In order to estimate the amplitude for this five-quark
configuration we use the cloudy bag model in which the pion-quark
coupling is dictated by chiral symmetry.
Motivated by the findings in the gluon case, we wish to check whether the
$\overline u$ antiquark is favoured over the $\overline d$
antiquark. If this turns out to be true, the asymmetry  
in favour of $\overline d$ (from the process $u\rightarrow d\pi^+$)
over $\overline u$ (from  $d\rightarrow u\pi^-$) may be at risk.

The relevant diagram to study antisymmetrization
in relation to the pion cloud is the one shown in
Fig. 3b, involving two loops. However, we will be mainly interested
in the relative importance of the two loop process in comparison with
that involving only 
one loop. To this end we also need to compute 
diagram (a) of Fig. \ref{figb4}, which we do as a warm-up exercise.
The calculations are going to be done at the quark level
($q\rightarrow \pi q$), which means 
that we need an interaction Hamiltonian between quarks and pions. 
To fix the calculation in a specific scheme, we shall use the 
interaction between quarks and pions as given by the 
Cloudy Bag Model \cite{theberge80,thomas84}, where the interaction is
totally specified by chiral symmetry:

\begin{figure}[htb]
\vspace{5cm}
\centering{\
    \epsfig{angle=270,figure=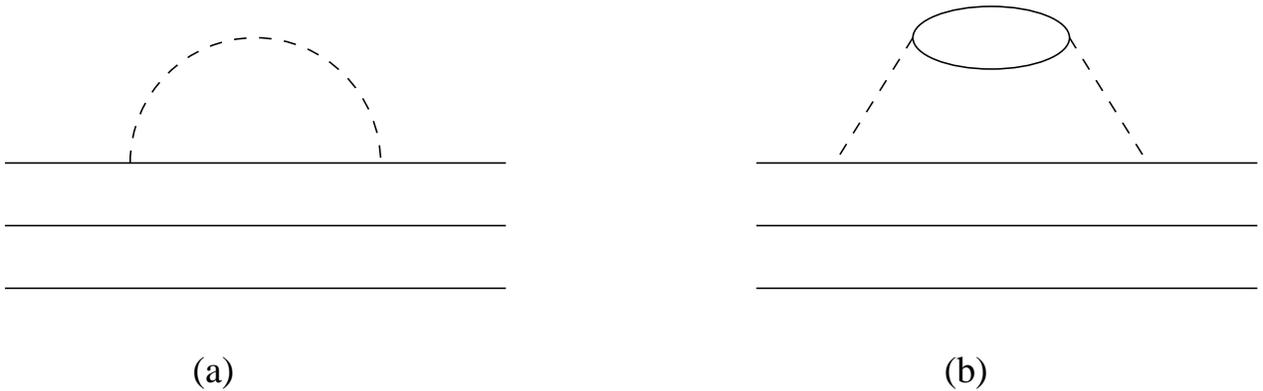,height=5cm}  }
\caption{The one and two loop graphs involved in pion emission.}
\label{figb4}
\end{figure}

\be
H_I = \frac{i}{2f}\int d^3 x \overline\psi (x) \gamma_5 
\tau^i \psi (x) \phi^i \delta(r - R),
\label{8}
\ee
with $R$ the bag radius and

\bear
\vec\phi (\vec x) &=& \frac{1}{(2\pi)^{2/3}}\int \frac{d\vec k}
{(2\omega_k)^{1/2}}(\vec a_{\vec k} e^{i\vec k \cdot \vec x} +
\vec a_{\vec k}^{\dagger} e^{-i\vec k \cdot \vec x}), \nonumber \\*
\psi^s (\vec x)&=&\frac{N}{(4\pi)^{1/2}}\sum\left\{
\left(\matrix{ij_0 (\omega r/R)\cr
              -j_1 (\omega r/R)\vec\sigma\cdot\hat r \cr}\right)\chi_m b(f,m,\alpha) +
\left(\matrix{-ij_1 (\omega r/R)\vec\sigma\cdot\hat r \cr
              j_0 (\omega r/R)\cr}\right)\chi_m d^{\dagger}(f,m,\alpha)\right\} \nonumber \\*
\psi^p (\vec x)&=&\frac{N}{(4\pi)^{1/2}}\sum\left\{
\left(\matrix{ij_1 (\omega r/R)\vec\sigma\cdot\hat r \cr
              j_0 (\omega r/R)\cr}\right)\chi_m b(f,m,\alpha) +
\left(\matrix{ij_0 (\omega r/R)\cr
              j_1 (\omega r/R)\vec\sigma\cdot\hat r \cr}\right)\chi_m d^{\dagger}(f,m,\alpha)\right\}.
\label{9}
\ear
Here $N^2=\omega^3/(2R^3 (\omega -1)sin^2 \omega)$, $\omega$ is the frequency
associated with a given principal quantum number and orbital angular momentum
and $j_0$ and $j_1$ are Bessel functions subject to the condition
$j_0 = j_1$ at the bag surface, $r=R$. We wrote the 
explicit forms for the $s$ and $p$ waves for a quark inside a cavity because 
they are going to be used later on. 

Three vertices are relevant to our diagrams: $q\rightarrow \pi q$,
$\pi\rightarrow q\overline q$ and $0\rightarrow \pi q\overline q$. 
As in the gluon case, there are some restrictions
due to the conservation of parity, that can be helpful. As is well known, 
for fermions a particle has
opposite intrinsic parity to the antiparticle. By convention, a particle has parity
$+ 1$ and an antiparticle parity $- 1$. Also, the parity of one particle 
relative to a set of others particles is given by $(-1)^l$, with $l$ the 
orbital angular momentum 
of the particle in question. If $\overline q$ has parity -1 then one 
of the quarks (or the same antiquark) has to be in a state $l=1$, or a p-wave, 
so that the proton parity is conserved. In general, the system must always be
in a state of even parity.

We now write down the explicit form for the interaction Hamiltonian for 
both vertices. It happens that their form, besides the 
creation operator for a quark or for a antiquark, is the same for both 
processes $q\rightarrow \pi q$ and $0\rightarrow \pi q\overline q$:

\bear
H_I^{q^{\tilde f} \rightarrow \pi q^f}&=& H_I^{0\rightarrow \pi q^f 
\overline{q}^{\tilde f}} =
\frac{i}{2f(2\pi)^{3/2}}\left(\frac{\omega^a \omega^b}{(\omega^a -1)
(\omega^b - 1)}\right) \nonumber \\*
&&\int \frac{d^3 k}{(2\omega_k)^{1/2}}\chi^{\dagger}_m \vec\sigma\cdot\vec k \chi_{\tilde m}
\frac{j_1 (kR)}{kR} b^{\dagger}(f,m,\alpha)\tau^i_{f\tilde f}
    \matrix{b(\tilde f, \tilde m, \tilde\alpha)\cr 
           d^{\dagger}(\tilde f, \tilde m, \tilde\alpha )\cr} a^{\dagger i}_{\vec k}.
\label{10}
\ear
For the case $\pi\rightarrow q\overline q$ the interaction is just
slightly different:

\bear
H_I^{\pi\rightarrow q^f \overline{q}^{\tilde f}} &=&
\frac{-i}{2f(2\pi)^{3/2}}\left(\frac{\omega^a \omega^b}{(\omega^a -1)
(\omega^b - 1)}\right) \nonumber \\*
&&\int \frac{d^3 k}{(2\omega_k)^{1/2}}\chi^{\dagger}_m \vec\sigma\cdot\vec k \chi_{\tilde m}
\frac{j_1 (kR)}{kR} b^{\dagger}(f,m,\alpha)\tau^i_{f\tilde f}
 d^{\dagger}(\tilde f, \tilde m, \tilde\alpha ) a^{i}_{\vec k}.
\label{101}
\ear
In expressions (\ref{10}) and (\ref{101}) the indices $a$ and $b$ 
refer to the spatial wavefunctions of the quarks, or/and antiquarks, 
involved in a specific reaction. 

Once we have the interaction Hamiltonian, it is easy to 
calculate the quantities in which we are interested. As announced, we
first calculate the probability to find a pion in the nucleon.
To be specific, we set $\omega =\omega^a = \omega^b$ in Eq. (\ref{10}),
which means that the quark remains in the ground state after it 
emits the pion. In this case we get:

\bear
\frac{1}{E_0 - H_0}H_I |p\rangle_0 &=& \frac{-i}{2f(2\pi)^{3/2}}\frac{\omega}{(\omega -1)}
\int \frac{d^3 k}{(2\omega_k)^{1/2}}\frac{1}{- \omega_k}
\chi^{\dagger}_m \vec\sigma\cdot\vec k \chi_{\tilde m}\frac{j_1 (kR)}{kR} \nonumber \\*
&&b^{\dagger}(v,m,\alpha)\tau^i_{v\tilde v}
      b(\tilde v, \tilde m, \tilde\alpha) a^{\dagger i}_{\vec k} |p\rangle_0 ,
\label{11}
\ear
where we used $(E_0 - H_0)(H_I |p\rangle_0) = (E_0 - (E_0 + \omega_k)) 
(H_I |p\rangle_0)$.
The probability to find a pion is then given by:

\bear
P_{\pi} &=& \left(_0\langle p|H_I\frac{1}{E_0 - H_0}\right)\left(\frac{1}
{E_0 - H_0}H_I |p\rangle_0\right) \nonumber \\*
&=&\frac{\pi}{6f^2(2\pi)^3}\frac{\omega^2}{(\omega - 1)^2}\int_0^{\infty}
\frac{k^4 dk}{\omega_k^3}\frac{j_1^2 (kR)}{(kR)^2} \nonumber \\*
&& \times _0\langle p|b^{\dagger}(\tilde{v}^{\prime},\tilde{m}^{\prime}
\tilde{\delta}^{\prime})
\tau^i_{\tilde{v}^{\prime}v^{\prime}}\sigma^j_{\tilde{m}^{\prime}m^{\prime}}
b(v^{\prime}, m^{\prime}, \delta^{\prime}) b^{\dagger}(v,m,\delta)\tau^i_{v\tilde v}
\sigma^j_{m\tilde m}b(\tilde v, \tilde m, \tilde\delta)|p\rangle_0 .
\label{12}
\ear
After some calculation, the expectation value between bare proton states is found
to be 57. In terms of the pion components, this result reads 57 =$22\delta_{\pi^+}
+ 19 \delta_{\pi^0} + 16\delta_{\pi^-}$. This means that, as expected, the
probability to find a $\overline d$ in the nucleon is bigger than to
find a $\overline u$. 
Using the identity $\omega/(f(\omega - 1)) = \sqrt{4\pi}18 f_{\pi NN}/
5m_{\pi}$, where $m_{\pi}$ is the physical pion mass and 
$f_{\pi NN}$ the pion nucleon nucleon coupling constant, expression (\ref{12}) 
is rewritten as:

\be
P_{\pi} = \frac{57}{25}\left(\frac{f_{\pi NN}}{m_{\pi}}\right)^2\frac{3}{\pi}
\int_0^{\infty} \frac{k^4 dk}{\omega_k^3}\left(\frac{3j_1 (kR)}{(kR)}\right)^2.
\label{13}
\ee

 
\begin{figure}[htb]
\vspace{5cm}
\centering{\
    \epsfig{angle=270,figure=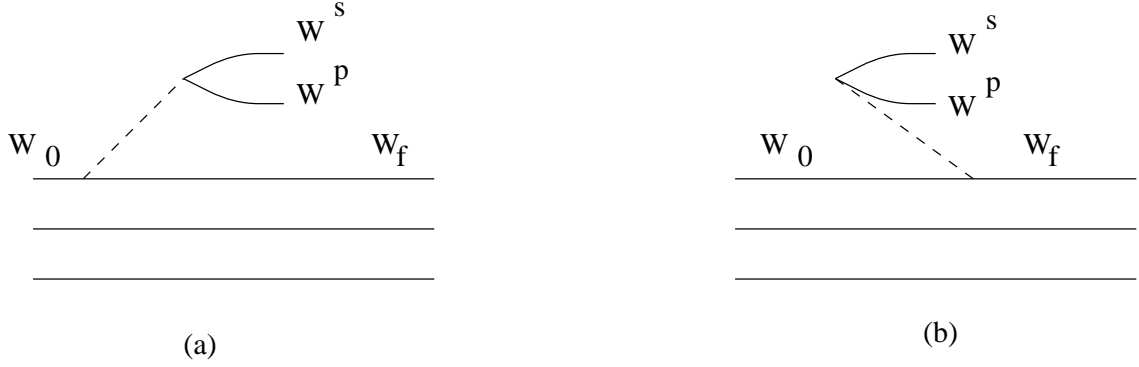,height=5cm}  }
\caption{The two loop graph in time order.}
\label{figb5}
\end{figure}

The next step is the evaluation of the two-loop graph. Its calculation
is a straightforward application of Eqs. (\ref{3}), (\ref{10}) and (\ref{101}).
We start with graph (a) of fig. \ref{figb5}. The process is pion creation
with subsequent decay into a $q\overline q$ pair:

\bear
\frac{1}{E_0 - H_0}H_I^{\pi\rightarrow q\overline q}\frac{1}{E_0 - H_0}H_I^{q\rightarrow\pi q}
|p\rangle_0 &=& \frac{1}{(2f)^2(2\pi)^3}\sqrt{\frac{\omega_f \omega_o \omega^s \omega^p}
{(\omega_f - 1)(\omega_o - 1)(\omega^s - 1)(\omega^p - 1)}} \nonumber \\*
   && \int d^3 k^{\prime} \int d^3 k \frac{1}{(2\omega_{k^{\prime}} 
2\omega_k)^{1/2}}
\frac{R}{\omega_o - \omega_f - R\omega_k}\frac{R}{\omega_o - \omega_f - 
\omega^s - \omega^p} \times \nonumber \\*
   &&\chi^{\dagger}_n \vec\sigma\cdot\vec{k}^{\prime} \chi_{\tilde n}
\frac{j_1 (k^{\prime} R)}
{k^{\prime}R}b^{\dagger}(s,n,\rho)\tau^j_{s\tilde s}
d^{\dagger}(\tilde s, \tilde n, \rho) a^j_{\vec{k}^{\prime}}\times \nonumber \\*
   &&\chi^{\dagger}_m \vec\sigma\cdot\vec k \chi_{\tilde m}\frac{j_1 (kR)}{kR} 
b^{\dagger}(v,m,\delta)\tau^i_{v\tilde v}
b(\tilde v, \tilde m, \delta) a^{\dagger i}_{\vec k} |p\rangle_0 .
\label{14}
\ear
The energies of the intermediate states are clearly indicated on the 
figure. The notation is the following: $\omega_0$ and $\omega_f$ are,
respectively, the quark frequencies before and after the pion emission 
or absorption, $\omega^s$ and $\omega^p$ are, respectively, the frequencies of the s wave
quark and of the $p$ wave antiquark. 

Similarly to case (a), we calculate the contribution from graph (b), where
a pion and a $q\overline q$ pair is created and the pion is 
subsequently absorbed by a quark:

\bear
\frac{1}{E_0 - H_0}H_I^{q\pi\rightarrow q}\frac{1}{E_0 - H_0}H_I^{0\rightarrow\pi q\overline q}
|p\rangle_0 &=& \frac{1}{(2f)^2(2\pi)^3}\sqrt{
\frac{\omega_f \omega_o \omega^s \omega^p}
{(\omega_f - 1)(\omega_o - 1)(\omega^s - 1)(\omega^p - 1)}} \nonumber \\*
   &&\int d^3 k^{\prime} \int d^3 k \frac{1}{(2\omega_{k^{\prime}} 2\omega_k)^{1/2}} 
\frac{R}{\omega_o - \omega_f -\omega^s - \omega^p}\frac{R}{-\omega^s - \omega^p - 
R\omega_k}\times \nonumber \\*
   &&\chi^{\dagger}_n \vec\sigma\cdot\vec{k}^{\prime} \chi_{\tilde n}
\frac{j_1 (k^{\prime} R)}
{k^{\prime}R}b^{\dagger}(v,n,\delta)\tau^i_{v\tilde v}
b^(\tilde v, \tilde n, \delta) a^i_{\vec{k}^{\prime}}\times \nonumber \\*
   &&\chi^{\dagger}_m \vec\sigma\cdot\vec k \chi_{\tilde m}\frac{j_1 (kR)}{kR} 
b^{\dagger}(s,m,\rho)\tau^j_{s\tilde s}
d^{\dagger}(\tilde s, \tilde m, \rho) a^{\dagger j}_{\vec k} |p\rangle_0 .
\label{15}
\ear
The total amplitude for the two loop process is then given by the sum of graphs (a) 
and (b). To better deal with this sum, we rewrite the product of operators in 
Eq. (\ref{15}) as:

\bear
&&b^{\dagger}(v,n,\delta)\tau^i_{v\tilde v}
b^(\tilde v, \tilde n, \delta) a^i_{\vec{k}^{\prime}}
b^{\dagger}(s,m,\rho)\tau^j_{s\tilde s}
d^{\dagger}(\tilde s, \tilde m, \rho) a^{\dagger j}_{\vec k} |p\rangle_0 \nonumber \\*
 &&= b^{\dagger}(s,n,\rho)\tau^j_{s\tilde s}d^{\dagger}(\tilde s, \tilde n, \rho)
b^{\dagger}(v,m,\delta)\tau^i_{v\tilde v}b(\tilde v, \tilde m, \delta)
a^i_{\vec{k}^{\prime}}a^{\dagger j}_{\vec k}|p\rangle_0 \nonumber \\*
 && + \delta_{\tilde v s}\delta_{\tilde n m}\delta_{\delta\rho}b^{\dagger}(v,n,\delta)
\tau^i_{v\tilde v}\tau^j_{s\tilde s}d^{\dagger}(\tilde s, \tilde m, \rho)
a^i_{\vec{k}^{\prime}}a^{\dagger j}_{\vec k})|p\rangle_0.
\label{16}
\ear
We then use $a^i_{\vec{k}^{\prime}}a^{\dagger j}_{\vec k}|0\rangle = \delta^{ij}
\delta(\vec{k}^{\prime} - \vec{k})|0\rangle$ to write the sum of the graphs (a) and (b)
as:

\bear
  &&\frac{1}{E_0 - H_0}H_I^{\pi\rightarrow q\overline q}\frac{1}{E_0 - H_0}H_I^{q\rightarrow\pi q}
|p\rangle_0 + \frac{1}{E_0 - H_0}H_I^{q\pi\rightarrow q}\frac{1}{E_0 - H_0}H_I^{0\rightarrow\pi q\overline q}
|p\rangle_0 \nonumber \\*
  &&=\frac{1}{(2f)^2(2\pi)^3}\sqrt{\frac{\omega_f \omega_o \omega^s \omega^p}
{(\omega_f - 1)(\omega_o - 1)(\omega^s - 1)(\omega^p - 1)}} \times \nonumber \\*
  &&\int d^3 k \frac{(\vec\sigma\cdot\vec k)_{n\tilde n}(\vec\sigma\cdot\vec k)_{m\tilde m}}
{2\omega_k}\frac{j_1^2 (kR)}{(kR)^2} \frac{R^2}{\omega_o - \omega_f - \omega^s - \omega^p}
\times \nonumber \\*
  &&\left\{
\left(\frac{1}{\omega_o - \omega_f - R\omega_k} + \frac{1}{-\omega^s - \omega^p - R\omega_k}\right)
b^{\dagger}(s,n,\rho)\tau^i_{s\tilde s}d^{\dagger}(\tilde s, \tilde n, \rho)
b^{\dagger}(v,m,\delta)\tau^i_{v\tilde v}b(\tilde v, \tilde m, \delta) \right. \nonumber \\*
  &&\left. + \frac{1}
{-\omega^s - \omega^p - R\omega_k}\delta_{\tilde v s}\delta_{\tilde n m}\delta_{\delta\rho}
b^{\dagger}(v,n,\delta)\tau^i_{v\tilde v}\tau^i_{s\tilde s}d^{\dagger}(\tilde s, \tilde m, \rho)
\right\}|p\rangle_0 .
\label{17}
\ear

We are now ready to compute the probability, $P_{\pi q\overline q}$, to find a 
$q \overline q$ pair in the nucleon. It is given by the square of the 
amplitude (\ref{17}):

\bear
P_{\pi q\overline q} &=& \left(\frac{1}{(2f)^2(2\pi)^3}\right)^2 \frac{\omega_f \omega_o \omega^s \omega^p}
{(\omega_f - 1)(\omega_o - 1)(\omega^s - 1)(\omega^p - 1)}\left(\frac{4\pi}{3}\right)^2 \nonumber \\*
  &&\int_o^{\infty} \frac{k^4 dk}{2\omega_k}\frac{j_1^2 (kR)}{(kR)^2}
\int_o^{\infty} \frac{k^{\prime 4} dk^{\prime}}{2\omega_{k^{\prime}}}
\frac{j_1^2 (k^{\prime}R)}{(k^{\prime}R)^2}\left(\frac{R^2}
{\omega_o - \omega_f - \omega^s - \omega^p}\right)^2 \times \nonumber \\*
  && \left\{\left(\frac{1}{\omega_o - \omega_f - R\omega_k} + 
\frac{1}{-\omega^s - \omega^p - R\omega_k}\right) \left(\frac{1}{\omega_o - \omega_f - 
R\omega_{k^{\prime}}} + \frac{1}{-\omega^s - \omega^p - R\omega_{k^{\prime}}}\right) \times 
\right. \nonumber \\*
  && \left. \hspace{1em} _0\langle p| b^{\dagger}(\tilde{v}^{\prime},\tilde{m}^{\prime},\delta^{\prime})
\tau^i_{\tilde{v}^{\prime} v^{\prime}} \sigma^j_{\tilde{m}^{\prime} m^{\prime}} 
b(v^{\prime}, m^{\prime}, \delta^{\prime})d(\tilde{s}^{\prime},\tilde{n}^{\prime},\rho^{\prime})
\tau^i_{\tilde{s}^{\prime} s^{\prime}} \sigma^j_{\tilde{n}^{\prime} n^{\prime}} 
b(v^{\prime}, m^{\prime}, \delta^{\prime}) \right. \nonumber \\*
  && \left. \hspace{7em}
b^{\dagger}(s,n,\rho)\tau^k_{s\tilde s}\sigma^l_{n\tilde n}d^{\dagger}(\tilde s, \tilde n, \rho)
b^{\dagger}(v,m,\delta)\tau^k_{v\tilde v}\sigma^l_{m\tilde m} b(\tilde v, \tilde m, \delta)|p\rangle_0
\right. \nonumber \\*
  && \left. \hspace{1em} + \frac{1}{-\omega^s - \omega^p - R\omega_k}\frac{1}{-\omega^s - \omega^p - 
R\omega_{k^{\prime}}}\times \right. \nonumber \\*
  &&\left. \hspace{1em} _0\langle p|d(\tilde{s}^{\prime}, \tilde{m}^{\prime}, \rho^{\prime})
\tau^i_{\tilde{s}^{\prime} s^{\prime}}\tau^i_{s^{\prime} v^{\prime}}
\sigma^j_{\tilde{m}^{\prime} m^{\prime}}\sigma^j_{m^{\prime} n^{\prime}}
b(v^{\prime}, n^{\prime}, \rho^{\prime})
b^{\dagger}(v,n,\rho)\tau^k_{vs}\tau^k_{s\tilde s}\sigma^l_{n m}
\sigma^l_{m \tilde m} d^{\dagger}(\tilde s, \tilde m, \rho) |p\rangle_0 \right\} .
\label{18}
\ear
The last quantities to be calculated are the expectation values between the bare 
proton states. 
It is a very long but straightforward calculation. We calculated only the case 
where the valence quark that emits or absorbs the pion remains in the same orbital state. 
The results are:

\bear
  && _0\langle p| b^{\dagger}(\tilde{v}^{\prime},\tilde{m}^{\prime},\delta^{\prime})
\tau^i_{\tilde{v}^{\prime} v^{\prime}} \sigma^j_{\tilde{m}^{\prime} m^{\prime}} 
b(v^{\prime}, m^{\prime}, \delta^{\prime})d(\tilde{s}^{\prime},\tilde{n}^{\prime},\rho^{\prime})
\tau^i_{\tilde{s}^{\prime} s^{\prime}} \sigma^j_{\tilde{n}^{\prime} n^{\prime}} 
b(v^{\prime}, m^{\prime}, \rho^{\prime}) \nonumber \\*
  && \hspace{6em}
b^{\dagger}(s,n,\rho)\tau^k_{s\tilde s}\sigma^l_{n\tilde n}d^{\dagger}(\tilde s, \tilde n, \rho)
b^{\dagger}(v,m,\delta)\tau^k_{v\tilde v}\sigma^l_{m\tilde m} b(\tilde v, \tilde m, \delta)|p\rangle_0
= \nonumber \\*
  && = 684 - 110\delta_{\tilde s \overline u} - 181 \delta_{\tilde s \overline d} \nonumber \\*
  && _0\langle p|d(\tilde{s}^{\prime}, \tilde{m}^{\prime}, \rho^{\prime})
\tau^i_{\tilde{s}^{\prime} s^{\prime}}\tau^i_{s^{\prime} v^{\prime}}
\sigma^j_{\tilde{m}^{\prime} m^{\prime}}\sigma^j_{m^{\prime} n^{\prime}}
b(v^{\prime}, n^{\prime}, \rho^{\prime})
b^{\dagger}(v,n,\rho)\tau^k_{vs}\tau^k_{s\tilde s}\sigma^l_{n m}
\sigma^l_{m \tilde m} d^{\dagger}(\tilde s, \tilde m, \rho) |p\rangle_0 \nonumber \\*
  && = 972 - 54\delta_{\tilde s \overline u} - 81 \delta_{\tilde s \overline d} .
\label{19}
\ear

The results contained in expressions (\ref{19}) are quite surprising. 
They say that, if the quark structure of the pion is important and if the
quarks from the pion are allowed to antisymmetrize with the
quarks from the parent proton, then the probability
for the antiquark in the pion to be a $\overline u$ is bigger than $\overline d$. 
The above result is independent of particularities of the given model, in the
sense that expressions (\ref{19}) are a direct consequence of the bare proton 
wave function, Eq. (\ref{1}). They are also a consequence 
of assuming a pion quark interaction. 

To better determine how important these second order effects could be, we will calculate 
the ratio of probabilities $P_{\pi q\overline q}/P_{\pi}$. To this end we need to 
perform the integrals over $k$ and $k^{\prime}$ in Eqs. (\ref{13}) and (\ref{18}).
These integrals are dependent on the particular value of the bag radius
and here we will display the results for $R=0.6,\; 0.8$ and $1\; fm$. We set
$\omega_0 = \omega_f$ and define the following integrals:

\bear
I_1&=& \frac{1}{2}\int_0^{\infty}\frac{dx}{m_{\pi}^2 R^2 + x^2}
\frac{\omega^s + \omega^p + 2(m_{\pi}^2 R^2 + x^2)^{1/2}}
{\omega^s + \omega^p + (m_{\pi}^2 R^2 + x^2)^{1/2}}
\left(\frac{sin^2 x}{x^2} + cos^2 x - \frac{sinx cosx}{x}\right),
\nonumber \\*
I_2&=& \frac{1}{2}\int_0^{\infty}\frac{dx}{(m_{\pi}^2 R^2 + x^2)^{1/2}}
\frac{1}{\omega^s + \omega^p + (m_{\pi}^2 R^2 + x^2)^{1/2}}
\left(\frac{sin^2 x}{x^2} + cos^2 x - \frac{sinx cosx}{x}\right),
\nonumber \\*
I_3&=& \int_0^{\infty}\frac{dx}{(m_{\pi}^2 R^2 + x^2)^{3/2}}
\left(\frac{sin^2 x}{x^2} + cos^2 x - \frac{sinx cosx}{x}\right),
\label{20}
\ear
where we use $x=kR$. The numerical values of these integrals are 
displayed in Table \ref{tb22}.
\newpage

\begin{table}
\begin{center}
\begin{tabular}{|c|c|c|c|}
\hline
Integral & $R=0.6\; fm$ & $R=0.8\; fm$ & $R=1\; fm$  \\ \hline
$I_1$ & 1.6848 & 1.24065 & 0.980306 \\
$I_2$ & 0.1946 & 0.1734430 & 0.157762 \\
$I_3$ & 4.92548 & 2.65299 & 1.63279 \\
$P_{\pi q\overline q}/P_{\pi}$ & 0.0217 & 0.0124 & 0.0081 \\
\hline
\end{tabular}
\end{center}
\caption{The two to one loop ratio for various bag radii.}
\label{tb22}
\end{table}
With these definitions, we rewrite expression
(\ref{18}) as:

\bear
P_{\pi q\overline q} &=& \left(\frac{1}{(2Rf)^2(2\pi)^3}\right)^2 \frac{\omega_o^2}
{(\omega_o - 1)^2}\frac{\omega^s \omega^p}
{(\omega^s - 1)(\omega^p - 1)}\left(\frac{4\pi}{3}\right)^2 
\left(\frac{1}{\omega^s + \omega^p}\right)^2\nonumber \\*
  &&\left\{I_1^2 \; (684 - 110\delta_{\tilde s \overline u} - 
181 \delta_{\tilde s \overline d}) + I_2^2 \; (972 - 54\delta_{\tilde\delta \overline u} - 
81 \delta_{\tilde\delta \overline d}) \right\}.
\label{21}
\ear
The ratio $P_{\pi q\overline q}/P_{\pi}$ is then easily expressed:

\bear
\frac{P_{\pi q\overline q}}{P_{\pi}}&=&
\frac{1}{(2Rf)^2(2\pi)^3} \frac{\omega^s \omega^p}
{(\omega^s - 1)(\omega^p - 1)}\frac{4\pi}{3} 
\left(\frac{1}{\omega^s + \omega^p}\right)^2 \nonumber \\*
  &&\left\{\frac{I_1^2}{I_3}\frac{(684 - 110\delta_{\tilde s \overline u} - 
181 \delta_{\tilde s \overline d})}{57} + \frac{I_2^2}{I_3} 
\frac{(972 - 54\delta_{\tilde\delta \overline u} - 
81 \delta_{\tilde\delta \overline d})}{57} \right\} .
\label{22}
\ear
The value of this ratio for different sizes of the bag is also displayed
in Table \ref{tb22}, where we used $m_{\pi} = 140\; MeV$ for the pion mass and
$f=93\; MeV$ for the pion decay constant. There is a strong dependence of
the calculated ratio on the bag size but, even for the worst scenario
(the case of a small bag), the size of the two-loop contribution is just
$2\%$ of the one loop term.

\section{Final Remarks}

Our primary interest has been to investigate the effect of anti-symmetry
between the valence quarks in the nucleon and the internal structure of
the pion. Within the cloudy bag model, in which the coupling of the pion
to quarks is dictated by chiral symmetry, we found that it
appears to be safe to neglect possible antisymmetrization effects
between pion quarks and the nucleon valence quarks. Of course, when writing
the nucleon wave function one would have to add the contributions from all
possible states:
\be
|N> = Z^{1/2} [|N>_0 + |N\pi>_0 + \sum_{q\; states}|N\pi q\overline q>_0 + ...],
\la{23}
\ee
where the sum of all quark states (1s, 2s, etc...) was particularly
emphasized. If the quark that emits the pion remains in the ground state
the only contribution from antisymmetry is the one in Table \ref{tb22}. As
we have to sum over all other possible states for this quark, it turns
out that the antisymmetry effects are further diluted. However, we
will also have some other contributions even in this case, because if the
quark in the pion is in the ground state it can antisymmetrize with the 
two spectator valence quarks. Also, if the quark in the pion is in an excited
state, it can combine with a particular excited state of the valence
quark that emitted the pion in the first place, as in the gluon example.
We did not make these calculations because of their level of complexity. 
Our goal
was to examine the behaviour of the dominant contribution, displayed
in Table \ref{tb22}, and our results indicate that this number is
itself already very small. 

Of course, as the sum over quark states in Eq. (\ref{23}) is infinite,
we can not rule out the possibility that eventually the antisymmetry 
graphs could be important. However, we believe that this is 
improbable, as the contributions from graphs which would be unaffected by
antisymmetry would grow even faster.
In summary, the major purpose of the present calculation was 
to check whether the antisymmetrization effects in pion emission can be safely
disregarded - they can.
\newline
\newline
\newline
We would like to thank to S. Gardner, D. Leinweber, W. Melnitchouk, K-F.
Liu, A. I. Signal and A. G. Williams for helpful discussions. 
This work was supported 
by the Australian Research Council and by CAPES (Brazil).

\addcontentsline{toc}{chapter}{\protect\numberline{}{References}}

\end{document}